\documentclass[twocolumn,amsmath,amssymb,aps,prl,nofootinbib,superscriptaddress,10pt]{revtex4-1}
\usepackage{graphicx}
\usepackage{braket}
\usepackage{amsmath}
\usepackage{microtype}
\usepackage{hyperref}
\usepackage{color}
\DeclareMathOperator{\Tr}{Tr}

\begin{document}

\title{Scale Invariant Entanglement Negativity at the Many-Body Localization Transition}
% Scale invariance at the Many-Body Localization Transition via Negativity?

\author{Johnnie Gray}
\affiliation{Department of Physics and Astronomy, University College London, London WC1E 6BT, United Kingdom}
\affiliation{QOLS, Blackett Laboratory, Imperial College London, London SW7 2AZ, United Kingdom}

\author{Abolfazl Bayat}
\affiliation{Institute of Fundamental and Frontier Sciences, University of Electronic Science and Technology of China, Chengdu 610051, China}
\affiliation{Department of Physics and Astronomy, University College London, London WC1E 6BT, United Kingdom}

\author{Arijeet Pal}
\affiliation{Department of Physics and Astronomy, University College London, London WC1E 6BT, United Kingdom}
\affiliation{London Centre for Nanotechnology, University College London, London WC1H 0AH, United Kingdom}

\author{Sougato Bose}
\affiliation{Department of Physics and Astronomy, University College London, London WC1E 6BT, United Kingdom}

\begin{abstract}
    The exact nature of the many-body localization transition remains an open question.
An aspect which has been posited in various studies is the emergence of scale invariance around this point, however the direct observation of this phenomenon is still absent.
    Here we achieve this by studying the logarithmic negativity and mutual information between disjoint blocks of varying size across the many-body localization transition.
    The two length scales, block sizes and the distance between them, provide a clear quantitative probe of scale invariance across different length scales.
    We find that at the transition point, the logarithmic negativity
    obeys a scale invariant exponential decay with respect to the ratio of block separation to size, whereas the mutual information obeys a polynomial decay. The observed scale invariance of the quantum correlations in a microscopic model opens the direction to probe the fractal structure in critical eigenstates using tensor network techniques and provide constraints on the theory of the many-body localization transition.
\end{abstract}

\maketitle

%\emph{Introduction.--}
%Many-body localized systems~\cite{anderson_absence_1958,basko_metalinsulator_2006,imbrie_diagonalization_2016} have become increasingly scrutinized both due to their characteristic lack of thermalization and the exotic nature of the transition between ergodic and localized phases which involves the full spectrum.
%The nature of this transition has in particular received much attention~\cite{pal_many-body_2010,luitz_many-body_2015,khemani_critical_2016} and motivated by standard quantum phase transitions, many works have shown finite size scaling and obtained critical exponents indicating second order type transition, although the issue is far from settled.
%The various quantities employed to do so include the energy level statistics, von Neumann entropy, and Schmidt gap~\cite{khemani_two_2017,gray2018many}.
%Nonetheless, there has been no direct observation of the scale invariant structure \emph{within} eigenstates rather than relying on varying the total system size.
%To address this, one needs to study quantities, unlike the above, which have a well defined and tunable scale -- necessarily bringing to mind mixed state measures where one has full control of how much of the system of consider.

%Why negativity: (i) MBL transition is characterized by the development of resonances in eigenstates which become rarer and more long-range eventually leading to an MBL phase.
Quantum entanglement has transformed our understanding of phases of matter and the transitions between them by revealing the complex quantum correlations in the states~\cite{sachdev2011quantum,raimond2001manipulating,osterloh2002scaling,osborne2002entanglement, de2012entanglement,bayat2012entanglement, alkurtass2016entanglement, wichterich2009scaling, marcovitch2009critical, vidal2003entanglement, its2005entanglement,calabrese2004entanglement, calabrese2013entanglement, calabrese2013entanglement, mbeng2017negativity, kitaev2006topological, bayat2014order}.
In certain topological and impurity models \cite{kitaev2006topological, bayat2014order}, tuning an external control parameter can result in novel ``entanglement phase transitions'' in which the entanglement in the ground state of the system reorganizes itself globally without any signatures in a local order parameter.
Recent developments in quantum phase transitions in highly excited states and non-equilibrium quantum orders~\cite{Huse2013LPQO,Pekker2014HG, bahri2015localization, Chandran2014SPT,  kjall2014many} highlights the importance of entanglement based understanding for theory and experiments~\cite{schreiber2015observation,choi2016exploring,luschen2017signatures,kohlert2019observation,smith2016many,xu2018emulating,ye2019propagation}.  This paradigm shift has been stimulated by the discovery of many-body localization (MBL)~\cite{basko_metalinsulator_2006, Oganesyan2007, pal2010many, nandkishore2015many, AbaninMBLReview}: the breakdown of thermalization in isolated, interacting, disordered quantum systems. The phase transition between MBL and thermal phases, usually tuned by the strength of disorder relative to interactions, is an entanglement phase transition where the entanglement structure of eigenstates and the entanglement dynamics undergo a singular change~\cite{pal2012thesis, Bauer2013, bardarson_unbounded_2012, serbyn2013universal, huse_phenomenology_2014, serbyn2013local,  nanduri2014entanglement, luitz2015many,  gray2018many, Yu2016, serbyn2016power, TomasiPRL2017, khemani2017critical}. However, the nature of the transition in microscopic models remains a fundamental open question.

%can the study of entanglement shed further light to the still mysterious nature of the thermal to MBL transition? For example whether such infinite temperature phase transitions show scale invariant behaviour similar to ground state criticality or finding out whether there is a bipartite/multipartite nature of the entanglement across the transition.

On the thermal side of the MBL transition, the entanglement entropy of the energy eigenstates satisfy a volume law, i.e. the entanglement entropy (EE) density is finite, consistent with the eigenstate thermalization hypothesis, while in the localized phase the eigenstates exhibit a boundary-law; EE density is zero.  This difference in scaling of entanglement entropy hints at a \textit{first order} character of the transition~\cite{Zhang2016, khemani_critical_2016, Goremykina2019, Dumitrescu2019}. On the other hand the divergence of the localization length on approaching the critical point from the localized side is reminiscent of a continuous, second-order phase transition~\cite{luitz2015many, Kulshreshtha2018, gray2018many, Herviou2019}. Furthermore, second order transitions are also characterized by scale invariance at the critical point. Many of these questions are hotly debated, and several questions about the transition remain unresolved as yet. For instance, how does scale invariance manifests itself in measures of entanglement across the MBL transition?

\begin{figure}[tb]
    \centering
    \includegraphics[width=\linewidth]{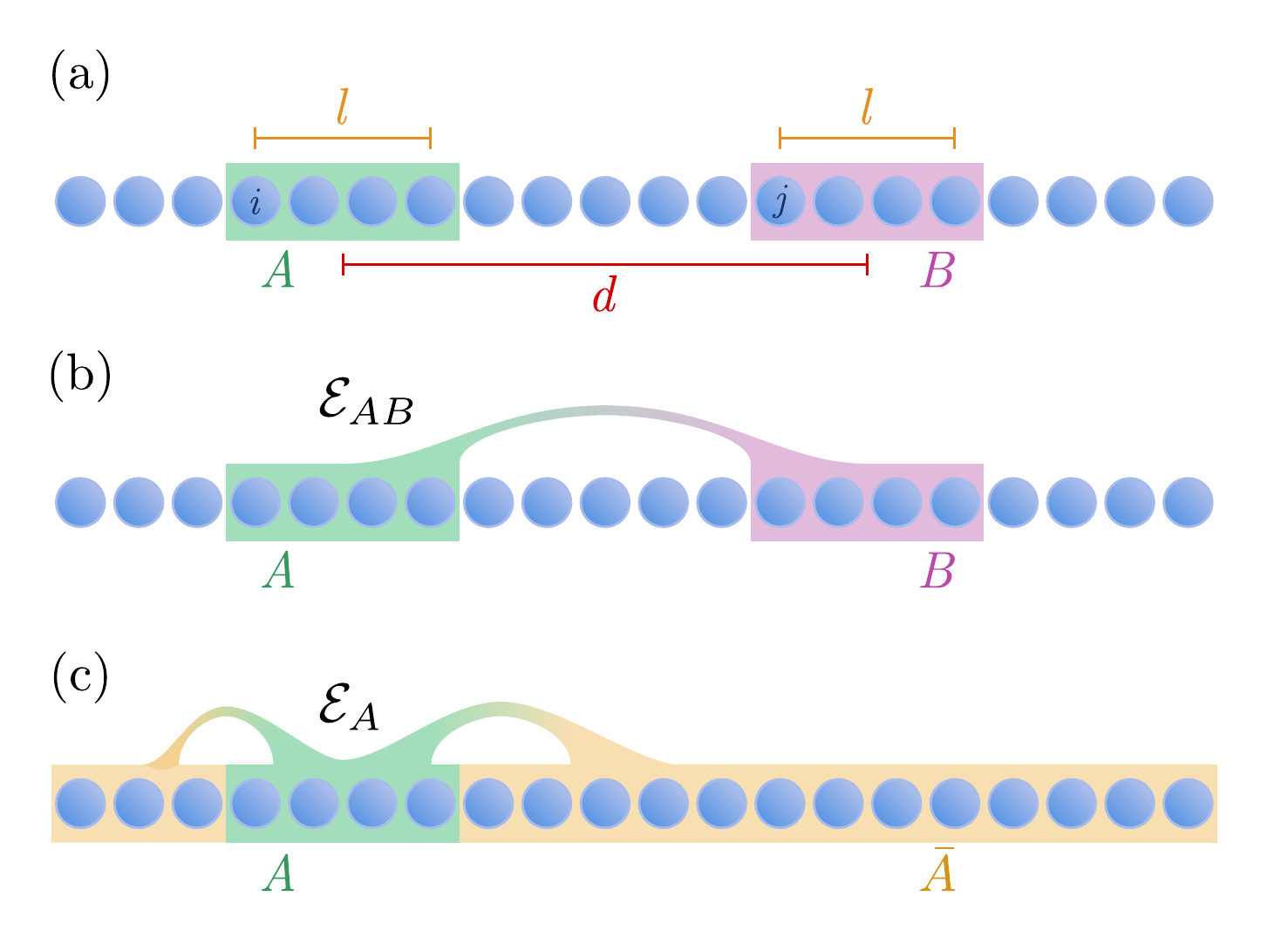}
    \caption{
    \textbf{Schematic of Set-Up.}
    (a) Definitions of the block size $l$, separation $d$ for disjoint blocks $A$ and $B$ located at sites $i$ and $j$ respectively of a spin chain of total length $L=20$.
    (b) The `bond' entanglement, as quantified by the logarithmic negativity $\mathcal{E}_{AB}$ between the two subsystems $A$ and $B$. The bond mutual information, $I_{AB}$, is similarly defined.
    (c) The `self' entanglement of subsystem $A$, which is simply the total entanglement between it and its complement (entire environment) $\bar{A}$.
    }
    \label{fig:schematic}
\end{figure}

\begin{figure*}[t!]
    \centering
    \includegraphics[width=\linewidth]{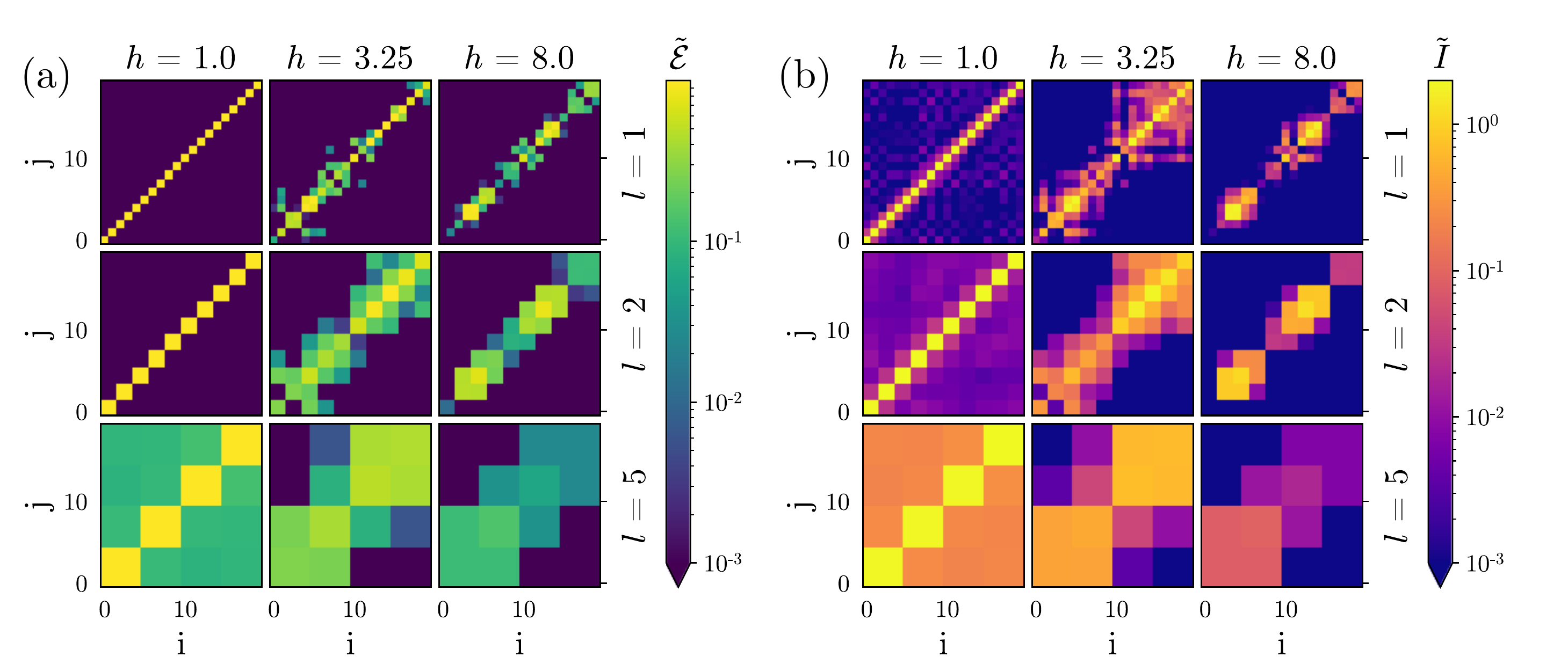}
    \caption{
    \textbf{Logarithmic Negativity \& Mutual Information Adjacency Matrices.}
    (a) Normalized self and bond logarithmic negativity between blocks beginning at $i$ and $j$ plotted as an adjacency-like matrix (where node weights are the self entanglement and edge weights are the bond entanglement) for various block sizes $l$ and random field strengths $h$.
    This data corresponds to a single noise field instance for $L=20$.
    (b) The same but substituting the normalized mutual information for logarithmic negativity.
    }
    \label{fig:raw_data_both}
\end{figure*}

In this article we unravel the scale invariant structure of entanglement at the MBL transition by focussing on the eigenstate logarithmic negativity (LN) ~\cite{zyczkowski1998volume,lee_partial_2000,vidal_computable_2002,plenio2005logarithmic}.
Unlike entanglement entropy and mutual information, LN is able to capture the quantum correlations even in mixed states, thus proving to be particularly informative for shedding light on the structure of many-body states~\cite{bayat2012entanglement, alkurtass2016entanglement, wichterich2009scaling,bayat2010negativity,gray2018many,coser2014entanglement,coser_towards_2016,santos2011negativity,PhysRevA.81.032311,bayat2017scaling}.
Additionally, by partitioning a system into non-complementary blocks one can extract information about the multipartite nature of entanglement in the system~\cite{bayat2017scaling}.
% thereby providing a distinct probe of the structure of entanglement in the vicinity of the MBL transition.
% Furthermore, LN captures the quantum part of the correlations whereas measures like mutual information receive contributions from classical correlations as well.
By investigating the LN across the MBL transition we reveal the emergence of scale-invariance in the quantum correlations at the transition point.
To our knowledge this is the first demonstration of such entanglement invariance, without assuming any prior finite size scaling ansatz, even in infinite temperature eigenstates.
% of entanglement across the whole eigenspectrum in the vicinity of a phase transition.
% This unequivocally shows that in the vicinity of the transition infinite temperature eigenstates display scale invariance.
% mid-spectrum eigenstates behave in a manner similar to ground states of hitherto studied quantum phase transitions.
We show striking differences between the LN and mutual information in the system revealing the multipartite  aspects of entanglement.
This scale invariance provides a constraint on phenomenological models developed for the MBL transition based on strong disorder renormalization group ~\cite{potter_universal_2015, Vosk2015, ZhangHuse2016, DumitrescuPRL2017, Goremykina2019, Dumitrescu2019, Morningstar2019}. For ground state infinite randomness fixed point, LN and MI scale identically which is distinct from the behaviour at the MBL transition found in this work~\cite{Ruggiero2016}. %Negativity has been extensively studied in many-body systems for characterising quantum phase transitions~\cite{bayat2012entanglement,alkurtass2016entanglement}, scaling properties in critical systems~\cite{wichterich2009scaling}, detecting entanglement length~\cite{bayat2010negativity,gray2018many} and capturing multipartite entanglement~\cite{bayat2017scaling}. This work is the first to explore the scale invariance at the MBL critical point using entanglement negativity.

\section{Results}

%\emph{Model.--}
\textbf{Model:}
We consider a spin-$1/2$ chain with random magnetic fields in the $z$-direction and with open boundary conditions:
\begin{equation}
    H = J \left(\sum_{i=1}^{L-1} \boldsymbol{S}_i \cdot \boldsymbol{S}_{i+1}
    +
    \sum_{i=1}^{L} h_i S^{z}_i \right)~,
\end{equation}
with $J$ the exchange coupling strength set to 1, $\boldsymbol{S}_i=\frac{1}{2} \left(\sigma^x_i, \sigma^y_i, \sigma^z_i\right)$ a vector of Pauli matrices acting on spin $i$ and dimensionless parameter $h_i$ the random field at site $i$ drawn from the uniform distribution $\left[-h, h\right]$.
For small $h$, this model exhibits thermalising behaviour, whereas for large $h$, it exhibits many-body localisation.
The transition point, $h_c$, between these two phases is suspected to lie between $h \sim 3.5-5$~\cite{luitz_many-body_2015,gray2018many}.
Unless otherwise noted we take $L=20$ (the longest numerically accessible size) and diagonalize the Hamiltonian in the spin-0 subspace.
% We diagonalize the Hamiltonian in either the spin-0 or spin-$\frac{1}{2}$ subspaces for even and odd $L$ respectively.
For each random instance a single eigenvector in the middle of the energy spectrum is evaluated~\cite{dalcin_parallel_2011,hernandez_slepc:_2005,gray2018quimb}.
%, half way between the minimum and maximum energies.
We compute relevant entanglement measures for these eigenstates, and unless explicitly noted average over many (at least 100) different noise realizations.

\begin{figure*}[t!]
    \centering
    \includegraphics[width=\linewidth]{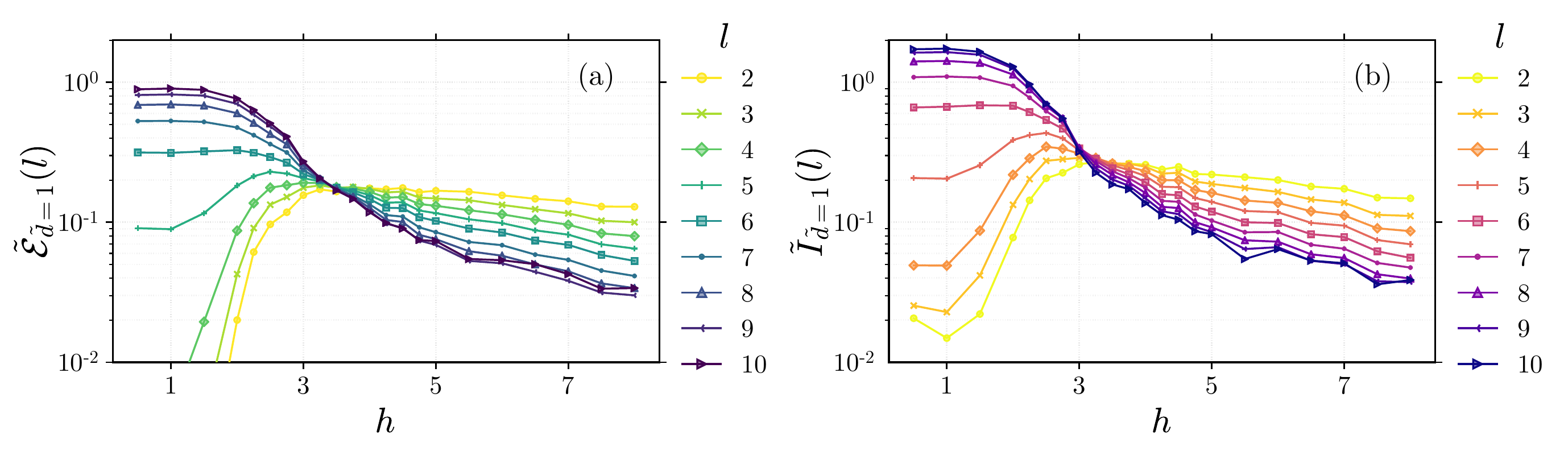}
    \caption{
    \textbf{Average Nearest Neighbour Blocked Logarithmic Negativity and Mutual Information.}
    (a) Total nearest neighbour logarithmic negativity, and (b) total nearest neighbour mutual information as a function of random field strength $h$ for various blocks sizes $l$.}
    \label{fig:logneg_mutinf_totalled}
\end{figure*}

\textbf{Multipartite entanglement:}
%In order to investigate scale invariance we need a probe with a well defined size -- in a specific sense.
% \textcolor{blue}{AP: Shorten this discussion so that we can get to the fig 2 in page 2}
In the MBL phase the  entanglement entropy in the eigenstates obeys the area-law. The local degrees of freedom become unentangled from their environment. In the vicinity of the transition the structure of the entanglement becomes more complex, where as on the other side of the transition in the thermal side, eigenstates satisfy the eigenstate thermalization hypothesis with the volume law for EE with the entanglement being dominated by highly non-local many-body degrees of freedom. We focus on two specific quantities to parse the structure of entanglement in the critical eigenstates, the logarithmic negativity~\cite{zyczkowski1998volume,lee_partial_2000,vidal_computable_2002,plenio2005logarithmic} and mutual information. These quantities are evaluated for two disjoint blocks (A and B) of equal size $l$, separated by a distance $d$, as in Fig.~\ref{fig:schematic}(a).
This allows us to vary both of these length scales whilst keeping their ratio fixed, a natural way to test for scale-invariance.
By contrast, the entanglement entropy of a single block is unsuitable, since growing the size of the block simultaneously shrinks its environment.
The two probes reveal complementary information about the state--  logarithmic negativity quantifies the \emph{quantum} correlations while the mutual information reflects the combined effect of classical and quantum correlations between the two subsystems.
The mutual information for a bipartite mixed state, $\rho_{AB}$, is defined as:
\begin{equation}
    I_{AB} = S_A + S_B - S_{AB}
\end{equation}
where $S_X$ denotes the von Neumann entropy of subsystem $X$ given by $S_X = -\Tr \left[ \rho_X\log_2 \rho_X \right]$.
This is bounded by $I_{AB} \leq 2l$.
The logarithmic negativity on the other hand is defined as
\begin{equation}
    \mathcal{E}_{AB} = \log_2 \left| \rho_{AB}^{T_B} \right|
\end{equation}
where $\cdot^{T_X}$ denotes the partial transpose with respect to subsystem $X$ and $|\cdot|$ the trace norm. We call this quantity,  which is bounded by $l$, the `bond' entanglement.
To access the logarithmic negativity when $2l > 12$ we use the TNSLQ method of~\cite{gray2018fast,gray2018quimb}.
Since we work with open boundary conditions, we can parametrize blocks $A$ and $B$ with coordinates $i$ and $j$ respectively such that the distance between their centres is $d = |i - j|$ as shown in Fig.~\ref{fig:schematic}(a)-(b).
Note that the smallest separation is thus $d=l$, corresponding to neighbouring blocks.
We also compute the logarithmic negativity for each single block with everything else (a quantity monotonically related to the entropy) which we call the `self' entanglement and denote as $\mathcal{E}_{A}=\mathcal{E}_{A\bar{A}}$, where $\bar{A}$ represents the complement of block $A$ -- as shown in Fig.~\ref{fig:schematic}(c).
Due to the monogamy of entanglement~\cite{coffman2000distributed} the self entanglement is always greater than or equal to the bond entanglement, namely $\mathcal{E}_A\ge \mathcal{E}_{AB}$.
This allows us to compare the portion of the self entanglement which is stored in bonds at a certain scale $l$
 % As an illustration, in the 3 qubit GHZ-state $(\ket{000} + \ket{111}) / \sqrt{2}$, each qubit is fully entangled with the remaining two, but no two qubits are specifically entangled with each other.
and thus infer how multipartite the entanglement is~\cite{adesso2007strong}.
We can also define the analogous quantity to the above by substituting the mutual information for the logarithmic negativity.
In this case the `self' mutual information, $I_A$, corresponds exactly to twice the von Neumann entropy of subsystem $A$.
Finally, in order to directly compare across different block sizes we also define the normalized entanglement negativity $\tilde{\mathcal{E}}=\mathcal{E}/l$, mutual information $\tilde{I}=I/l$, and distance $\tilde{d}=d/l$.

\begin{figure*}[t!]
    \centering
    \includegraphics[width=\linewidth]{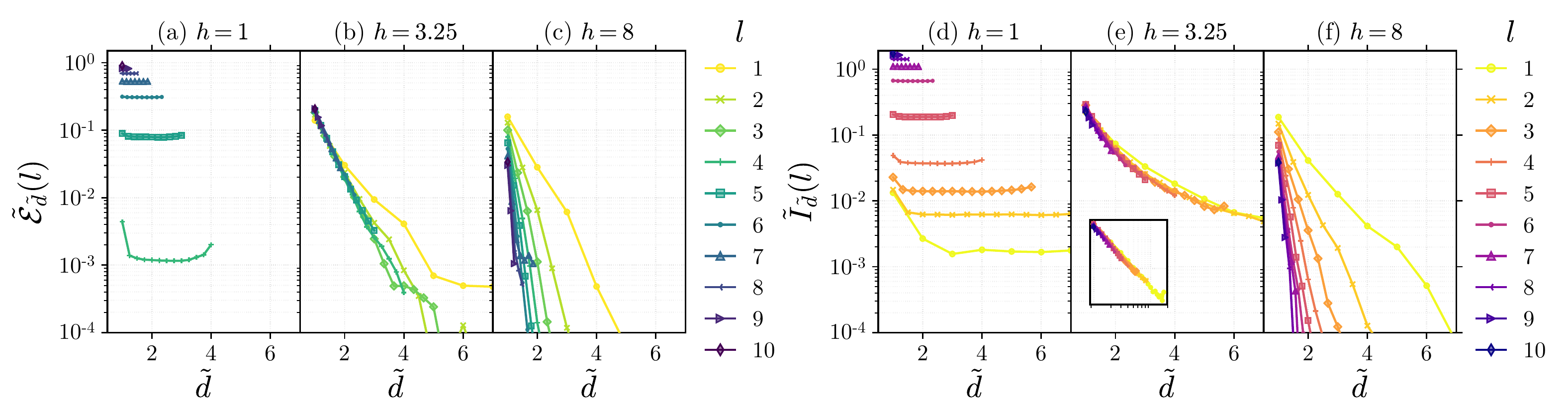}
    \caption{
    \textbf{Decay of bond logarithmic negativity and mutual information with relative separation.}
    Decay of normalized logarithmic negativity (panels (a)-(c)) and mutual information (panels (d)-(f)) with normalized separation $\tilde{d}$ for several random field strengths $h$ and various block sizes $l$ for $L=20$.
    (a) and (d) correspond to the ergodic regime with $h=1$.
    (b) and (e) show data near the transition point with $h \sim h_c \sim 3.25$.
    The inset of panel (e) is the same data, but plotted on a log-log scale in order to demonstrate the power law behaviour.
    A stretched exponential fit was also investigated but this was not found to be as natural as a power law.
    (c) and (f) correspond to the MBL regime.
    }
    \label{fig:logneg_gap_raw_decay}
\end{figure*}

In Fig.~\ref{fig:raw_data_both}(a) we show the data for both normalized self and bond entanglement as a matrix with elements $(i,j)$ for varying block size $l$ and a  single instance each for random field strength $h$.
This is analogous to an adjacency matrix where the node weights are the self entanglement and the edge weights the bond entanglement.
The diagonal terms, i.e. $i=j$, represent the self entanglement of the block $i$ while the off diagonal elements denote the bond entanglement between blocks $i$ and $j$.
As the figure shows in the ergodic regime for any block size $l$ and location $i$ the self entanglement is almost maximum (i.e. $\tilde{\mathcal{E}}_A \sim 1$).
% \textcolor{blue}{AP: Clarify this discussion}
On the other hand the bond entanglement is zero due to the multipartite nature of entanglement -- until the size of $A \cup B$ becomes half of the system size, namely $l \sim L/4$.
This matches the expected behaviour for random pure states~\cite{bhosale2012entanglement,gray2018fast}.
Near the transition point, however, bond entanglement appears for all scales of block size $l$ including rare cases where $d$ is many multiples of $l$.
We associate these to the emergence of resonances between distant blocks in the eigenstates of the system.
We also note that while certain spins show no individual ($l=1$) bond entanglement, by increasing the block size to $l=2$ for example, entanglement is revealed between the same \emph{groups} of spins, implying the multipartite nature of that entanglement.
By increasing $h$ into the MBL phase, the bond entanglement becomes short range and all the entanglement structure is apparent in block size $l=1$, whereas large blocks have relatively diminished entanglement.
This can be explained as in the MBL phase the entanglement is mainly area-law like and thus primarily within the blocks rather than between them.

In Fig.~\ref{fig:raw_data_both}(b) we show the equivalent plots for the normalized mutual information.
Although many of the same structural patterns appear, there is a major difference with the logarithmic negativity.
Namely, in the ergodic regime and close to the transition, there is mutual information between individual spins and small blocks where no entanglement is detected.
This further evidences multipartite entanglement, the imprint of which
% , in loose analogy to the GHZ-state,
is solely classical correlations among subsystems.
In the MBL regime we also note the presence of mutual information at longer ranges than the logarithmic negativity.

% \begin{figure}[tb]
%     \centering
%     \includegraphics[width=\linewidth]{logneg_mutinf_totalled}
%     \caption{
%     \textbf{Total Self and Nearest Neighbour Logarithmic Negativity and Mutual Information.}
%     (a) Total self logarithmic negativity, (b) total nearest neighbour logarithmic negativity, (c) total self mutual information, and (d) total nearest neighbour mutual information as a function of random field strength $h$ for various blocks sizes $l$.}
%     \label{fig:logneg_mutinf_totalled}
% \end{figure}

Generally speaking the LN and MI depend on both $l$ and $d$, but for scale invariance, we would expect instead only the ratio of $l$ and $d$ to be relevant.
As such, we now introduce two quantities that depend on block size, $l$, and \emph{normalized} separation $\tilde{d} = d / l$.
From these, scale invariance would be heralded by the disappearance of any dependence on $l$.
Specifically, we consider the relative LN and MI averaged over all pairs of blocks with the same relative separation $\tilde{d}$:
\begin{align}
    \tilde{\mathcal{E}}_{\tilde{d}}(l)
    &=
    \frac{1}{\mathcal{N}}
    \sum_{\{i, j:{~}\frac{|i-j|}{l}=\tilde{d}\}}
    \left \langle
    \tilde{\mathcal{E}}_{A_i B_j}
    \right \rangle \cr
    \tilde{I}_{\tilde{d}}(l)
    &=
    \frac{1}{\mathcal{N}}
    \sum_{\{i, j:{~}\frac{|i-j|}{l}=\tilde{d}\}}
    \left \langle
    \tilde{I}_{A_i B_j}
    \right \rangle~.
\end{align}
where $\mathcal{N}$ is the size of the summation set and $\langle \cdot \rangle$ indicates the ensemble average over noise instances.
For example, we can consider only the entanglement or correlations contained in nearest neighbour blocks, namely, $\tilde{d}=1$.
In Fig.~\ref{fig:logneg_mutinf_totalled}(a) we show $\tilde{\mathcal{E}}_{\tilde{d}=1}(l)$ as a function of $h$ across the transition for varying block size $l$.
At $h \sim 3.25$ we find a very clear data collapse for all curves -- in other words, the dependence on block-size $l$ drops out.
We infer that $h \sim 3.25$ corresponds to the transition point for this total system size of $L=20$, which matches previous studies~\cite{gray2018fast}.
We find very similar behaviour for the equivalent MI quantity, $\tilde{I}_{\tilde{d}=1}(l)$ in Fig.~\ref{fig:logneg_mutinf_totalled}(b).
We emphasize that this is really quite distinct with respect to previous studies, where the collapse has been with regard to total system size $L$~\cite{luitz_many-body_2015,khemani_critical_2016,gray2018many}.

To probe the scale invariance in even stronger terms we can consider not only the nearest neighbour entanglement but the full behaviour with regard to arbitrary block separation.
% Specifically, we now consider the relative entanglement averaged over all pairs of blocks with the same relative separation $\tilde{d}$:
% \begin{align}
%     \mathcal{E}_{\tilde{d}}(l)
%     &=
%     \frac{1}{\mathcal{N}}
%     \sum_{\{i, j:{~}\frac{|i-j|}{l}=\tilde{d}\}}
%     \left \langle
%     \tilde{\mathcal{E}}_{A_i B_j}
%     \right \rangle \cr
%     I_{\tilde{d}}(l)
%     &=
%     \frac{1}{\mathcal{N}}
%     \sum_{\{i, j:{~}\frac{|i-j|}{l}=\tilde{d}\}}
%     \left \langle
%     \tilde{I}_{A_i B_j}
%     \right \rangle~.
% \end{align}
% where $\mathcal{N}$ is the size of the summation set.
% % We equivalently define $I_{\tilde{d}}(l)$.
We plot $\tilde{\mathcal{E}}_{\tilde{d}}(l)$ and $\tilde{I}_{\tilde{d}}(l)$ as functions of $\tilde{d}$ with varying $l$ in Fig.~\ref{fig:logneg_gap_raw_decay} for three representative values of $h$ and a total length of $L=20$.
In Fig.~\ref{fig:logneg_gap_raw_decay}(a), deep in the ergodic phase, we find that for blocks large enough to have bond entanglement ($l \gtrsim L / 4$), there is essentially no dependence on the separation $\tilde{d}$ -- the states are permutationally invariant as expected for volume law entanglement.
In Fig.~\ref{fig:logneg_gap_raw_decay}(b), at approximately the transition point for this length, $h_c \sim 3.25$, all the curves collapse onto each other with an exponential decay - there is no dependence on $l$.
Finally, in Fig.~\ref{fig:logneg_gap_raw_decay}(c), deep in the MBL phase, the entanglement $\mathcal{E}_{\tilde{d}}(l)$ depends on both $l$ and $\tilde{d}$ and decays quicker relative to larger $l$ -- as expected for area law states.

In Figs.~\ref{fig:logneg_gap_raw_decay}(d)-(f), we plot
$I_{\tilde{d}}(l)$
for the same choices of $h$.
Although qualitatively the behaviour is similar -- constant in the ergodic phase, collapse near the transition point and fast decay in the MBL phase -- there are a few notable differences.
Firstly, the mutual information is much more pervasive that the logarithmic negativity -- classical correlations appear for small blocks in the ergodic phase at further separations in the MBL phase.
Moreover, at the transition point, the behaviour observed is a collapse to power law decay rather than exponential decay, as demonstrated by the log-log inset plot of Fig.~\ref{fig:logneg_gap_raw_decay}(e).

% \begin{figure}[tb]
%     \centering
%     \includegraphics[width=\linewidth]{logneg_mutinf_decay_fit}
%     \caption{
%     \textbf{Collapse of Fitted Ansatz Parameters.}
%     Parameters fitted via least squares regression as functions of random strength $h$ and block size $l$ for the logarithmic negativity ansatz in Eq.~\eqref{eq:ansatz_decay_logneg} (panels (a) and (c)) and the mutual information ansatz in Eq.~\eqref{eq:ansatz_decay_mutinf} (panels (b) and (d)).
%     Total system size is $L=20$.
%     }
%     \label{fig:logneg_mutinf_decay_fit}
% \end{figure}

The full data collapse in Fig.~\ref{fig:logneg_gap_raw_decay}(b) and (e) is strong evidence for scale invariance at the transition point.
Based on this we suggest the following ansatzes for the decay of the logarithmic negativity and mutual information near the transition point:
\begin{align}
    \mathcal{E}_{\tilde{d}}(l)
    &= \label{eq:ansatz_decay_logneg}
    C_{\mathcal{E}} e^{-\tilde{d} / \lambda_{\mathcal{E}}}
    \\
    I_{\tilde{d}}(l)
    &= \label{eq:ansatz_decay_mutinf}
    C_{I} \tilde{d}^{-1 / \alpha_I}
\end{align}
where generically $C_{\mathcal{E}}$, $C_{I}$, $\lambda_{\mathcal{E}}$ and $\alpha_I$ might all be functions of $L$, $l$ and $h$.
To investigate these quantities we perform least squares fitting of Eq.~\eqref{eq:ansatz_decay_logneg} and Eq.~\eqref{eq:ansatz_decay_mutinf} to our data in the region of the transition,
% In Figs.~\ref{fig:logneg_mutinf_decay_fit}(a)-(d) we plot the coefficients found as a function of $h$ and $l$ and indeed find that at the critical point the dependence on $l$ for all coefficients disappears.
finding the values of the coefficients to be
$C_{\mathcal{E}} \sim 2$,
$\lambda_{\mathcal{E}} \sim 0.45$,
$C_{I} \sim 0.3$, and
$\alpha_I \sim 0.5$.

\begin{figure}[t!]
    \centering
    \includegraphics[width=\linewidth]{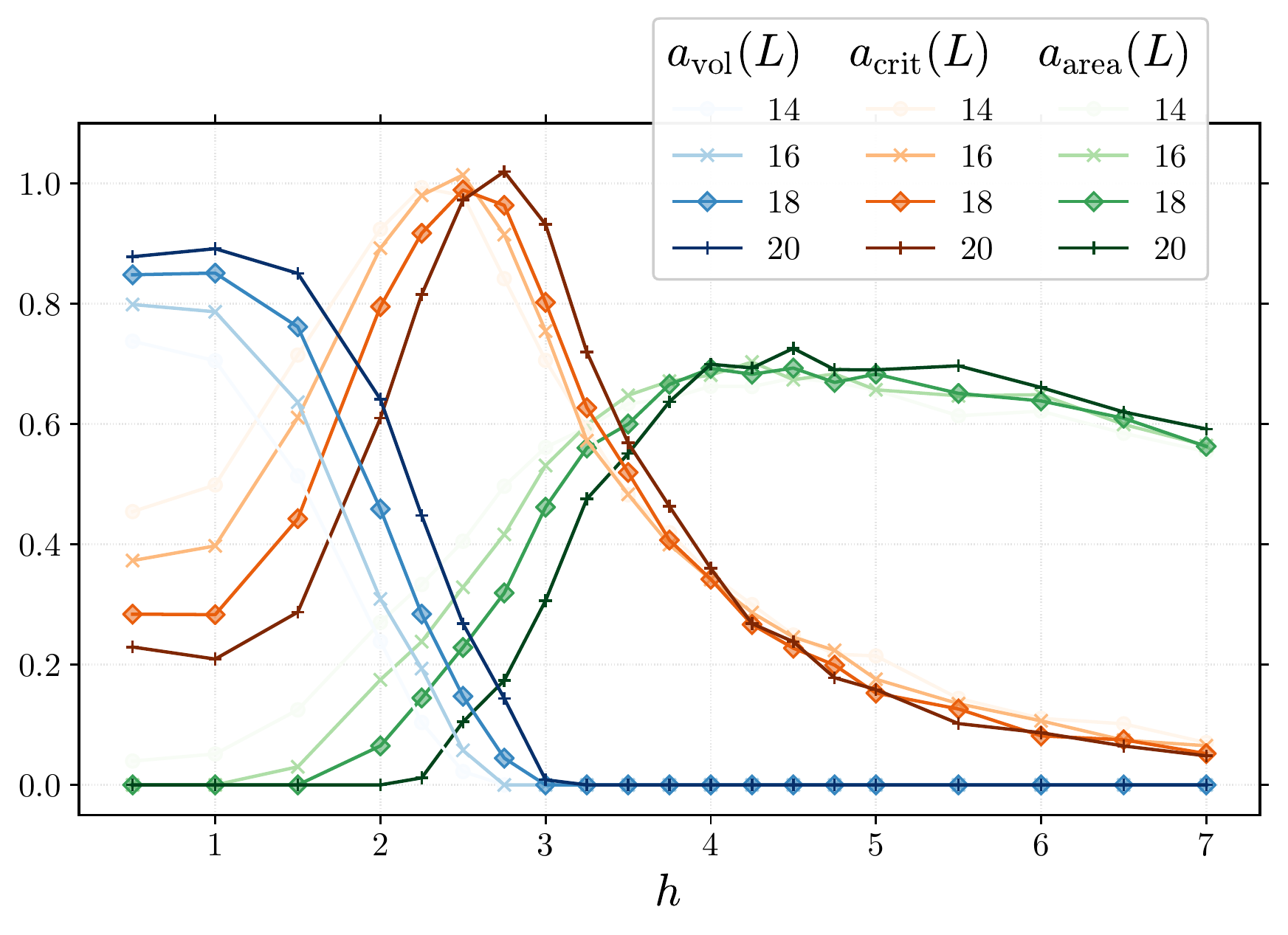}
    \caption{
        \textbf{Emergence of $\log l$ scaling of the self entanglement at the MBL transition.}
        Coefficients extracted from least-squares fitting the average self-entanglement as $\langle \mathcal{E}_A (l) \rangle = a_\mathrm{vol} l + a_\mathrm{crit} \log_2{l} + a_\mathrm{area}$ for varying random field strength $h$ and total system size $L$.
        The mean uncertainty derived from the fitting process for the three coefficients was 0.057, 0.16, and 0.094 respectively.
    }
    \label{fig:self_ent_fitting}
\end{figure}

\textbf{$\mathbf{\mathbf\mathrm{log}(l)}$ scaling of self entanglement:}
Our above analysis of the microscopic entanglement structure reveals strong evidence of scale invariance.
One interesting parallel to draw here are low energy states of disorder-free gap-less models, for which scale invariance is captured by logarithmic scaling of the entanglement entropy with block size - directly related to the self entanglement, $\mathcal{E}_A$ studied here.
To further investigate this link we fit an ansatz to the average (taken here over both random samples and block locations) self entanglement of the form
\begin{equation}
   \langle \mathcal{E}_A(l) \rangle = a_\mathrm{vol} l + a_\mathrm{crit} \log_2{l} + a_\mathrm{area}
\end{equation}
for varying block size $l$, total length $L$ and random field strength $h$.
In Fig~\ref{fig:self_ent_fitting} we plot the extracted coefficients $a_\mathrm{vol}, a_\mathrm{crit}, a_\mathrm{area}$ using least-squares fitting including their uncertainty.
We exclude the smallest and largest block size $l=\{1, L / 2\}$ to mitigate finite size effects.
As the figure shows, in the ergodic regime, the behaviour is well described by a linear scaling of $\mathcal{E}_A$ with $l$ -- volume law behaviour with $a_\mathrm{vol}$ approaching $1$ whilst $ a_\mathrm{crit}$ and $a_\mathrm{area}$ both approach 0.
Deep in the MBL regime, we find the behaviour is dominated by the constant term, $a_\mathrm{area}$, reflecting an area law as expected.
Near the transition however, we indeed find that both $a_\mathrm{vol}$ and $a_\mathrm{area}$ are small and instead the logarithmic scaling term $a_\mathrm{crit}$ peaks.
Both the ergodic and critical features described can be seen to sharpen with increasing total system size $L$, whilst the MBL behaviour is, as expected, fixed.
These features further corroborate the scale invariant structure of entanglement in critical eigenstates at the transition point.

\section{Discussion}
% \textbf{Conclusions.--}

The nature of the MBL transition is still not well-understood, although the emergence of scale invariance near the transition point has already been conjectured.
Here however we have directly observed the scale invariant structure \emph{within} states around the MBL transition.
We have done so by employing the logarithmic negativity and mutual information, both of which permit two controllable lengths -- block size $l$ and separation $d$.
We found that at the transition point, the `total' amount of entanglement and mutual information stored in nearest neighbour ($d=l$) bonds collapses for all block sizes $l$.
Motivated by this, we investigated the decay of the normalized logarithmic and mutual information as functions of normalized separation $\tilde{d}$, finding collapse across all block sizes for both.
Crucially, for the entanglement we find exponential decay whereas for the full correlations, as quantified by the mutual information, we find polynomial decay.
The final piece of evidence provided is that the average self entanglement $\langle \mathcal{E}_A \rangle$ is well described by a logarithmic scaling with block size $l$ close to the transition, as opposed to linear and constant scaling in the ergodic and MBL regimes respectively.

One immediate implication of this work is the likely existence of a Multiscale Entanglement Renormalization Ansatz (MERA)~\cite{VidalMERA, EvenblyMERA} like description of critical eigenstates near the MBL transition. Such a description would extend the efficient simulations from the fully MBL regime~\cite{yu2015finding, Khemani2016MPS, Chandran2015STN, Pollmann2016TNS, Pekker2017MPO, Wahl2017, wahl2D_2019} into the critical region, an area currently limited essentially to exact diagonalization and thus small system sizes. This can potentially open directions for analysing infinite temperature eigenstate quantum phase transitions using controlled numerical techniques. Furthermore, the behaviour of entanglement negativity across the MBL transition provides a constraint on the theory of the MBL transition using strong disorder renormalization group of highly excited states.

\section*{Acknowledgements}

A. P. acknowledges helpful conversations with Jens Bardarson, Bryan Clark, Tarun Grover, and  Shivaji Sondhi.
J. G. acknowledges funding from the EPSRC Centre for Doctoral Training in Delivering Quantum Technologies at UCL and the Samsung Advanced Institute of Technology Global Research Partnership.
S. B. and A. B. thank EPSRC grant EP/R029075/1 (Non-Ergodic Quantum Manipulation).
A. B. also thanks the National Key R\&D Program of China, Grant No. 2018YFA0306703.

\bibliography{mblvr-main}{}

\begin{thebibliography}{10}

\bibitem{sachdev2011quantum}
S.~Sachdev, {\em Quantum phase transitions}.
\newblock Cambridge University Press, 2011.

\bibitem{raimond2001manipulating}
J.-M. Raimond, M.~Brune, and S.~Haroche, ``Manipulating quantum entanglement
  with atoms and photons in a cavity,'' {\em Rev. Mod. Phys.}, vol.~73, no.~3,
  p.~565, 2001.

\bibitem{osterloh2002scaling}
A.~Osterloh, L.~Amico, G.~Falci, and R.~Fazio, ``Scaling of entanglement close
  to a quantum phase transition,'' {\em Nature}, vol.~416, no.~6881, p.~608,
  2002.

\bibitem{osborne2002entanglement}
T.~J. Osborne and M.~A. Nielsen, ``Entanglement in a simple quantum phase
  transition,'' {\em Phys. Rev. A}, vol.~66, no.~3, p.~032110, 2002.

\bibitem{de2012entanglement}
G.~De~Chiara, L.~Lepori, M.~Lewenstein, and A.~Sanpera, ``Entanglement
  spectrum, critical exponents, and order parameters in quantum spin chains,''
  {\em Phys. Rev. Lett.}, vol.~109, no.~23, p.~237208, 2012.

\bibitem{bayat2012entanglement}
A.~Bayat, S.~Bose, P.~Sodano, and H.~Johannesson, ``Entanglement probe of
  two-impurity kondo physics in a spin chain,'' {\em Phys. Rev. Lett.},
  vol.~109, no.~6, p.~066403, 2012.

\bibitem{alkurtass2016entanglement}
B.~Alkurtass, A.~Bayat, I.~Affleck, S.~Bose, H.~Johannesson, P.~Sodano, E.~S.
  S{\o}rensen, and K.~Le~Hur, ``Entanglement structure of the two-channel kondo
  model,'' {\em Phys. Rev. B}, vol.~93, no.~8, p.~081106, 2016.

\bibitem{wichterich2009scaling}
H.~Wichterich, J.~Molina-Vilaplana, and S.~Bose, ``Scaling of entanglement
  between separated blocks in spin chains at criticality,'' {\em Phys. Rev. A},
  vol.~80, no.~1, p.~010304, 2009.

\bibitem{marcovitch2009critical}
S.~Marcovitch, A.~Retzker, M.~Plenio, and B.~Reznik, ``Critical and noncritical
  long-range entanglement in klein-gordon fields,'' {\em Phys. Rev. A},
  vol.~80, no.~1, p.~012325, 2009.

\bibitem{vidal2003entanglement}
G.~Vidal, J.~I. Latorre, E.~Rico, and A.~Kitaev, ``Entanglement in quantum
  critical phenomena,'' {\em Phys. Rev. Lett.}, vol.~90, no.~22, p.~227902,
  2003.

\bibitem{its2005entanglement}
A.~R. Its, B.-Q. Jin, and V.~E. Korepin, ``Entanglement in the xy spin chain,''
  {\em Journal of Physics A: Mathematical and General}, vol.~38, no.~13,
  p.~2975, 2005.

\bibitem{calabrese2004entanglement}
P.~Calabrese and J.~Cardy, ``Entanglement entropy and quantum field theory,''
  {\em Journal of Statistical Mechanics: Theory and Experiment}, vol.~2004,
  no.~06, p.~P06002, 2004.

\bibitem{calabrese2013entanglement}
P.~Calabrese, J.~Cardy, and E.~Tonni, ``Entanglement negativity in extended
  systems: a field theoretical approach,'' {\em Journal of Statistical
  Mechanics: Theory and Experiment}, vol.~2013, no.~02, p.~P02008, 2013.

\bibitem{mbeng2017negativity}
G.~B. Mbeng, V.~Alba, and P.~Calabrese, ``Negativity spectrum in 1d gapped
  phases of matter,'' {\em Journal of Physics A: Mathematical and Theoretical},
  vol.~50, no.~19, p.~194001, 2017.

\bibitem{kitaev2006topological}
A.~Kitaev and J.~Preskill, ``Topological entanglement entropy,'' {\em Phys.
  Rev. Lett.}, vol.~96, no.~11, p.~110404, 2006.

\bibitem{bayat2014order}
A.~Bayat, H.~Johannesson, S.~Bose, and P.~Sodano, ``An order parameter for
  impurity systems at quantum criticality,'' {\em Nat. Commun.}, vol.~5,
  p.~3784, 2014.

\bibitem{Huse2013LPQO}
D.~A. Huse, R.~Nandkishore, V.~Oganesyan, A.~Pal, and S.~L. Sondhi,
  ``Localization-protected quantum order,'' {\em Phys. Rev. B}, vol.~88,
  p.~014206, Jul 2013.

\bibitem{Pekker2014HG}
D.~Pekker, G.~Refael, E.~Altman, E.~Demler, and V.~Oganesyan, ``{Hilbert-Glass
  Transition: New Universality of Temperature-Tuned Many-Body Dynamical Quantum
  Criticality},'' {\em Phys. Rev. X}, vol.~4, p.~011052, Mar 2014.

\bibitem{bahri2015localization}
Y.~Bahri, R.~Vosk, E.~Altman, and A.~Vishwanath, ``{Localization and topology
  protected quantum coherence at the edge of hot matter},'' {\em Nature
  communications}, vol.~6, p.~7341, 2015.

\bibitem{Chandran2014SPT}
A.~Chandran, V.~Khemani, C.~R. Laumann, and S.~L. Sondhi, ``Many-body
  localization and symmetry-protected topological order,'' {\em Phys. Rev. B},
  vol.~89, p.~144201, Apr 2014.

\bibitem{kjall2014many}
J.~A. Kj{\"a}ll, J.~H. Bardarson, and F.~Pollmann, ``{Many-body localization in
  a disordered quantum Ising chain},'' {\em Phys. Rev. Lett.}, vol.~113,
  no.~10, p.~107204, 2014.

\bibitem{schreiber2015observation}
M.~Schreiber, S.~S. Hodgman, P.~Bordia, H.~P. L{\"u}schen, M.~H. Fischer,
  R.~Vosk, E.~Altman, U.~Schneider, and I.~Bloch, ``Observation of many-body
  localization of interacting fermions in a quasirandom optical lattice,'' {\em
  Science}, vol.~349, no.~6250, pp.~842--845, 2015.

\bibitem{choi2016exploring}
J.-y. Choi, S.~Hild, J.~Zeiher, P.~Schau{\ss}, A.~Rubio-Abadal, T.~Yefsah,
  V.~Khemani, D.~A. Huse, I.~Bloch, and C.~Gross, ``Exploring the many-body
  localization transition in two dimensions,'' {\em Science}, vol.~352,
  no.~6293, pp.~1547--1552, 2016.

\bibitem{luschen2017signatures}
H.~P. L{\"u}schen, P.~Bordia, S.~S. Hodgman, M.~Schreiber, S.~Sarkar, A.~J.
  Daley, M.~H. Fischer, E.~Altman, I.~Bloch, and U.~Schneider, ``Signatures of
  many-body localization in a controlled open quantum system,'' {\em Phys. Rev.
  X}, vol.~7, no.~1, p.~011034, 2017.

\bibitem{kohlert2019observation}
T.~Kohlert, S.~Scherg, X.~Li, H.~P. L{\"u}schen, S.~D. Sarma, I.~Bloch, and
  M.~Aidelsburger, ``Observation of many-body localization in a one-dimensional
  system with a single-particle mobility edge,'' {\em Phys. Rev. Lett.},
  vol.~122, no.~17, p.~170403, 2019.

\bibitem{smith2016many}
J.~Smith, A.~Lee, P.~Richerme, B.~Neyenhuis, P.~W. Hess, P.~Hauke, M.~Heyl,
  D.~A. Huse, and C.~Monroe, ``Many-body localization in a quantum simulator
  with programmable random disorder,'' {\em Nat. Phys.}, vol.~12, no.~10,
  p.~907, 2016.

\bibitem{xu2018emulating}
K.~Xu, J.-J. Chen, Y.~Zeng, Y.-R. Zhang, C.~Song, W.~Liu, Q.~Guo, P.~Zhang,
  D.~Xu, H.~Deng, {\em et~al.}, ``Emulating many-body localization with a
  superconducting quantum processor,'' {\em Phys. Rev. Lett.}, vol.~120, no.~5,
  p.~050507, 2018.

\bibitem{ye2019propagation}
Y.~Ye, Z.-Y. Ge, Y.~Wu, S.~Wang, M.~Gong, Y.-R. Zhang, Q.~Zhu, R.~Yang, S.~Li,
  F.~Liang, {\em et~al.}, ``Propagation and localization of collective
  excitations on a 24-qubit superconducting processor,'' {\em Phys. Rev.
  Lett.}, vol.~123, no.~5, p.~050502, 2019.

\bibitem{basko_metalinsulator_2006}
D.~M. Basko, I.~L. Aleiner, and B.~L. Altshuler, ``Metal\textendash{}insulator
  transition in a weakly interacting many-electron system with localized
  single-particle states,'' {\em Ann. Phys.}, vol.~321, pp.~1126--1205, May
  2006.

\bibitem{Oganesyan2007}
V.~Oganesyan and D.~A. Huse, ``{Localization of interacting fermions at high
  temperature},'' {\em Phys. Rev. B}, vol.~75, no.~15, p.~155111, 2007.

\bibitem{pal2010many}
A.~Pal and D.~A. Huse, ``Many-body localization phase transition,'' {\em Phys.
  Rev. B}, vol.~82, no.~17, p.~174411, 2010.

\bibitem{nandkishore2015many}
R.~Nandkishore and D.~A. Huse, ``{Many-Body Localization and Thermalization in
  Quantum Statistical Mechanics},'' {\em Annual Review of Condensed Matter
  Physics}, vol.~6, pp.~15--38, 2015.

\bibitem{AbaninMBLReview}
D.~A. Abanin, E.~Altman, I.~Bloch, and M.~Serbyn, ``Colloquium: Many-body
  localization, thermalization, and entanglement,'' {\em Rev. Mod. Phys.},
  vol.~91, p.~021001, May 2019.

\bibitem{pal2012thesis}
A.~Pal, ``Many-body localization,'' 2012.

\bibitem{Bauer2013}
B.~Bauer and C.~Nayak, ``{Area laws in a many-body localized state and its
  implications for topological order},'' {\em Journal Of Statistical
  Mechanics-Theory And Experiment}, vol.~2013, p.~P09005, sep 2013.

\bibitem{bardarson_unbounded_2012}
J.~H. Bardarson, F.~Pollmann, and J.~E. Moore, ``Unbounded {{Growth}} of
  {{Entanglement}} in {{Models}} of {{Many}}-{{Body Localization}},'' {\em
  Phys. Rev. Lett.}, vol.~109, p.~017202, July 2012.

\bibitem{serbyn2013universal}
M.~Serbyn, Z.~Papi{\'c}, and D.~A. Abanin, ``Universal slow growth of
  entanglement in interacting strongly disordered systems,'' {\em Phys. Rev.
  Lett.}, vol.~110, no.~26, p.~260601, 2013.

\bibitem{huse_phenomenology_2014}
D.~A. Huse, R.~Nandkishore, and V.~Oganesyan, ``Phenomenology of fully
  many-body-localized systems,'' {\em Phys. Rev. B}, vol.~90, p.~174202, Nov.
  2014.

\bibitem{serbyn2013local}
M.~Serbyn, Z.~Papi{\'c}, and D.~A. Abanin, ``Local conservation laws and the
  structure of the many-body localized states,'' {\em Phys.l Rev. Lett.},
  vol.~111, no.~12, p.~127201, 2013.

\bibitem{nanduri2014entanglement}
A.~Nanduri, H.~Kim, and D.~A. Huse, ``Entanglement spreading in a many-body
  localized system,'' {\em Phys. Rev. B}, vol.~90, no.~6, p.~064201, 2014.

\bibitem{luitz2015many}
D.~J. Luitz, N.~Laflorencie, and F.~Alet, ``Many-body localization edge in the
  random-field heisenberg chain,'' {\em Phys. Rev. B}, vol.~91, no.~8,
  p.~081103, 2015.

\bibitem{gray2018many}
J.~Gray, S.~Bose, and A.~Bayat, ``Many-body localization transition: Schmidt
  gap, entanglement length, and scaling,'' {\em Phys. Rev. B}, vol.~97, no.~20,
  p.~201105, 2018.

\bibitem{Yu2016}
X.~Yu, D.~J. Luitz, and B.~K. Clark, ``Bimodal entanglement entropy
  distribution in the many-body localization transition,'' {\em Phys. Rev. B},
  vol.~94, p.~184202, Nov 2016.

\bibitem{serbyn2016power}
M.~Serbyn, A.~A. Michailidis, D.~A. Abanin, and Z.~Papi{\'c}, ``Power-law
  entanglement spectrum in many-body localized phases,'' {\em Phys. Rev.
  Lett.}, vol.~117, no.~16, p.~160601, 2016.

\bibitem{TomasiPRL2017}
G.~De~Tomasi, S.~Bera, J.~H. Bardarson, and F.~Pollmann, ``Quantum mutual
  information as a probe for many-body localization,'' {\em Phys. Rev. Lett.},
  vol.~118, p.~016804, Jan 2017.

\bibitem{khemani2017critical}
V.~Khemani, S.-P. Lim, D.~Sheng, and D.~A. Huse, ``Critical properties of the
  many-body localization transition,'' {\em Phys. Rev. X}, vol.~7, no.~2,
  p.~021013, 2017.

\bibitem{Zhang2016}
L.~Zhang, B.~Zhao, T.~Devakul, and D.~A. Huse, ``Many-body localization phase
  transition: A simplified strong-randomness approximate renormalization
  group,'' {\em Phys. Rev. B}, vol.~93, p.~224201, Jun 2016.

\bibitem{khemani_critical_2016}
V.~Khemani, S.~P. Lim, D.~N. Sheng, and D.~A. Huse, ``Critical {{Properties}}
  of the {{Many}}-{{Body Localization Transition}},'' {\em arXiv:1607.05756},
  July 2016.

\bibitem{Goremykina2019}
A.~Goremykina, R.~Vasseur, and M.~Serbyn, ``Analytically solvable
  renormalization group for the many-body localization transition,'' {\em Phys.
  Rev. Lett.}, vol.~122, p.~040601, Jan 2019.

\bibitem{Dumitrescu2019}
P.~T. Dumitrescu, A.~Goremykina, S.~A. Parameswaran, M.~Serbyn, and R.~Vasseur,
  ``Kosterlitz-thouless scaling at many-body localization phase transitions,''
  {\em Phys. Rev. B}, vol.~99, p.~094205, Mar 2019.

\bibitem{Kulshreshtha2018}
A.~K. Kulshreshtha, A.~Pal, T.~B. Wahl, and S.~H. Simon, ``Behavior of l-bits
  near the many-body localization transition,'' {\em Phys. Rev. B}, vol.~98,
  p.~184201, Nov 2018.

\bibitem{Herviou2019}
L.~Herviou, S.~Bera, and J.~H. Bardarson, ``Multiscale entanglement clusters at
  the many-body localization phase transition,'' {\em Phys. Rev. B}, vol.~99,
  p.~134205, Apr 2019.

\bibitem{zyczkowski1998volume}
K.~{\.Z}yczkowski, P.~Horodecki, A.~Sanpera, and M.~Lewenstein, ``Volume of the
  set of separable states,'' {\em Phys. Rev. A}, vol.~58, no.~2, p.~883, 1998.

\bibitem{lee_partial_2000}
J.~Lee, M.~Kim, Y.~Park, and S.~Lee, ``Partial teleportation of entanglement in
  a noisy environment,'' {\em J. Mod. Opt.}, vol.~47, pp.~2151--2164, Oct.
  2000.

\bibitem{vidal_computable_2002}
G.~Vidal and R.~F. Werner, ``Computable measure of entanglement,'' {\em Phys.
  Rev. A}, vol.~65, p.~032314, Feb. 2002.

\bibitem{plenio2005logarithmic}
M.~B. Plenio, ``Logarithmic negativity: a full entanglement monotone that is
  not convex,'' {\em Phys. Rev. Lett.}, vol.~95, no.~9, p.~090503, 2005.

\bibitem{bayat2010negativity}
A.~Bayat, P.~Sodano, and S.~Bose, ``Negativity as the entanglement measure to
  probe the kondo regime in the spin-chain kondo model,'' {\em Physical Review
  B}, vol.~81, no.~6, p.~064429, 2010.

\bibitem{coser2014entanglement}
A.~Coser, E.~Tonni, and P.~Calabrese, ``Entanglement negativity after a global
  quantum quench,'' {\em Journal of Statistical Mechanics: Theory and
  Experiment}, vol.~2014, no.~12, p.~P12017, 2014.

\bibitem{coser_towards_2016}
A.~Coser, E.~Tonni, and P.~Calabrese, ``Towards the entanglement negativity of
  two disjoint intervals for a one dimensional free fermion,'' {\em Journal of
  Statistical Mechanics: Theory and Experiment}, vol.~2016, no.~3, p.~033116,
  2016.

\bibitem{santos2011negativity}
R.~A. Santos, V.~Korepin, and S.~Bose, ``Negativity for two blocks in the
  one-dimensional spin-1 affleck-kennedy-lieb-tasaki model,'' {\em Phys. Rev.
  A}, vol.~84, no.~6, p.~062307, 2011.

\bibitem{PhysRevA.81.032311}
H.~Wichterich, J.~Vidal, and S.~Bose, ``Universality of the negativity in the
  lipkin-meshkov-glick model,'' {\em Phys. Rev. A}, vol.~81, p.~032311, Mar
  2010.

\bibitem{bayat2017scaling}
A.~Bayat, ``Scaling of tripartite entanglement at impurity quantum phase
  transitions,'' {\em Phys. Rev. Lett.}, vol.~118, no.~3, p.~036102, 2017.

\bibitem{potter_universal_2015}
A.~C. Potter, R.~Vasseur, and A.~Parameswaran, S., ``Universal {{Properties}}
  of {{Many}}-{{Body Delocalization Transitions}},'' {\em Phys. Rev. X},
  vol.~5, p.~031033, Sept. 2015.

\bibitem{Vosk2015}
R.~Vosk, D.~A. Huse, and E.~Altman, ``{Theory of the Many-Body Localization
  Transition in One-Dimensional Systems},'' {\em Phys. Rev. X}, vol.~5,
  p.~31032, sep 2015.

\bibitem{ZhangHuse2016}
L.~Zhang, B.~Zhao, T.~Devakul, and D.~A. Huse, ``Many-body localization phase
  transition: A simplified strong-randomness approximate renormalization
  group,'' {\em Phys. Rev. B}, vol.~93, p.~224201, Jun 2016.

\bibitem{DumitrescuPRL2017}
P.~T. Dumitrescu, R.~Vasseur, and A.~C. Potter, ``Scaling theory of
  entanglement at the many-body localization transition,'' {\em Phys. Rev.
  Lett.}, vol.~119, p.~110604, Sep 2017.

\bibitem{Morningstar2019}
A.~Morningstar and D.~A. Huse, ``Renormalization-group study of the many-body
  localization transition in one dimension,'' {\em Phys. Rev. B}, vol.~99,
  p.~224205, Jun 2019.

\bibitem{Ruggiero2016}
P.~Ruggiero, V.~Alba, and P.~Calabrese, ``Entanglement negativity in random
  spin chains,'' {\em Phys. Rev. B}, vol.~94, p.~035152, Jul 2016.

\bibitem{luitz_many-body_2015}
D.~J. Luitz, N.~Laflorencie, and F.~Alet, ``Many-body localization edge in the
  random-field {{Heisenberg}} chain,'' {\em Phys. Rev. B}, vol.~91, p.~081103,
  Feb. 2015.

\bibitem{dalcin_parallel_2011}
L.~D. Dalcin, R.~R. Paz, P.~A. Kler, and A.~Cosimo, ``Parallel distributed
  computing using {{Python}},'' {\em Adv. Water Resour.}, vol.~34,
  pp.~1124--1139, Sept. 2011.

\bibitem{hernandez_slepc:_2005}
V.~Hernandez, J.~E. Roman, and V.~Vidal, ``{{SLEPc}}: {{A Scalable}} and
  {{Flexible Toolkit}} for the {{Solution}} of {{Eigenvalue Problems}},'' {\em
  ACM Trans. Math. Softw.}, vol.~31, pp.~351--362, Sept. 2005.

\bibitem{gray2018quimb}
J.~Gray, ``\texttt{quimb}: a python library for quantum information and
  many-body calculations,'' {\em Journal of Open Source Software}, vol.~3,
  no.~29, p.~819, 2018.

\bibitem{gray2018fast}
J.~Gray, ``Fast computation of many-body entanglement,'' {\em arXiv preprint
  arXiv:1809.01685}, 2018.

\bibitem{coffman2000distributed}
V.~Coffman, J.~Kundu, and W.~K. Wootters, ``Distributed entanglement,'' {\em
  Physical Review A}, vol.~61, no.~5, p.~052306, 2000.

\bibitem{adesso2007strong}
G.~Adesso and F.~Illuminati, ``Strong monogamy of bipartite and genuine
  multipartite entanglement: the gaussian case,'' {\em Phys. Rev. Lett.},
  vol.~99, no.~15, p.~150501, 2007.

\bibitem{bhosale2012entanglement}
U.~T. Bhosale, S.~Tomsovic, and A.~Lakshminarayan, ``Entanglement between two
  subsystems, the wigner semicircle and extreme-value statistics,'' {\em Phys.
  Rev. A}, vol.~85, no.~6, p.~062331, 2012.

\bibitem{VidalMERA}
G.~Vidal, ``Class of quantum many-body states that can be efficiently
  simulated,'' {\em Phys. Rev. Lett.}, vol.~101, p.~110501, Sep 2008.

\bibitem{EvenblyMERA}
G.~Evenbly and G.~Vidal, ``Algorithms for entanglement renormalization,'' {\em
  Phys. Rev. B}, vol.~79, p.~144108, Apr 2009.

\bibitem{yu2015finding}
X.~Yu, D.~Pekker, and B.~K. Clark, ``{Finding Matrix Product State
  Representations of Highly Excited Eigenstates of Many-Body Localized
  Hamiltonians},'' {\em Phys. Rev. Lett.}, vol.~118, p.~017201, Jan 2017.

\bibitem{Khemani2016MPS}
V.~Khemani, F.~Pollmann, and S.~L. Sondhi, ``{Obtaining Highly Excited
  Eigenstates of Many-Body Localized Hamiltonians by the Density Matrix
  Renormalization Group Approach},'' {\em Phys. Rev. Lett.}, vol.~116,
  p.~247204, Jun 2016.

\bibitem{Chandran2015STN}
A.~Chandran, J.~Carrasquilla, I.~H. Kim, D.~A. Abanin, and G.~Vidal,
  ``{Spectral tensor networks for many-body localization},'' {\em Phys. Rev.
  B}, vol.~92, p.~024201, Jul 2015.

\bibitem{Pollmann2016TNS}
F.~Pollmann, V.~Khemani, J.~I. Cirac, and S.~L. Sondhi, ``{Efficient
  variational diagonalization of fully many-body localized Hamiltonians},''
  {\em Phys. Rev. B}, vol.~94, p.~041116, Jul 2016.

\bibitem{Pekker2017MPO}
D.~Pekker and B.~K. Clark, ``Encoding the structure of many-body localization
  with matrix product operators,'' {\em Phys. Rev. B}, vol.~95, p.~035116, Jan
  2017.

\bibitem{Wahl2017}
T.~B. Wahl, A.~Pal, and S.~H. Simon, ``Efficient representation of fully
  many-body localized systems using tensor networks,'' {\em Phys. Rev. X},
  vol.~7, p.~021018, May 2017.

\bibitem{wahl2D_2019}
T.~B. Wahl, A.~Pal, and S.~H. Simon, ``Signatures of the many-body localized
  regime in two dimensions,'' {\em Nature Physics}, vol.~15, no.~2, p.~164,
  2019.

\end{thebibliography}
\bibliographystyle{ieeetr}

\end{document}